\documentclass[12pt,twoside]{article}
\usepackage{amssymb}
\def\beq{\begin{equation}}
\def\eeq{\end{equation}}
\def\bea{\begin{eqnarray}}
\def\eea{\end{eqnarray}}
\def\ba{\begin{array}}
\def\ea{\end{array}}
\def\nn{\nonumber} 
\def\lrb{\left(}
\def\rrb{\right)} 
\def\lsb{\left[}
\def\rsb{\right]} 
\def\lcb{\left\{}
\def\rcb{\right\}}
\def\fr{\frac}

\def\lra{\longrightarrow}
\def\h{\hat} 
\def\lb{\label}
\def\s{\sigma} 
\def\o{\omega}
\def\ox{\otimes}
\def\vf{\varphi}
\def\O{\Omega} 
\def\e{\epsilon} 
\def\t{\tau} 
\def\T{\cal T} 
\def\u{\underline}
\def\l{\lambda} 
\def\L{\Lambda} 
\def\a{\alpha}
\def\b{\beta}
\def\cL{\cal L} 
\def\p{\partial} 
\def\mq{{\mbox q}}
\def\mi{{\mbox i}}
\def\mj{{\mbox j}}
\def\mk{{\mbox k}} 
\def\v{\vec} 
\def\lra{\longrightarrow} 
\begin{document}  
\begin{center}
{\Large\bf  On generalized Clifford algebras \\ 
and their physical applications} 
\footnote{{\em AMS Classification Numbers}\,: 15A66, 20C25, 20C35, 81R05.} \\ 
\bigskip
R. Jagannathan\footnote{Formerly of MATSCIENCE, The Institute of Mathematical Sciences, Chennai. \\ 
{\em Email}: {\tt jagan@cmi.ac.in, jagan@imsc.res.in}}\\
\smallskip 
{\em Chennai Mathematical Institute \\ 
Plot H1, SIPCOT IT Park, Padur P.O. \\ 
Siruseri 603103, Tamilnadu, India} 
\end{center} 
\medskip 
\centerline{\sf Dedicated to the memory of Professor Alladi Ramakrishnan} 
\vspace{1cm}
\begin{abstract}
Generalized Clifford algebras (GCAs) and their physical applications were extensively studied 
for about a decade from 1967 by Alladi Ramakrishnan and his collaborators under the name of 
$L$-matrix theory.   Some aspects of GCAs and their physical applications are outlined here.   
The topics dealt with include: GCAs and projective representations of finite abelian groups, 
Alladi Ramakrishnan's $\s$-operation approach to the representation theory of Clifford algebra 
and GCAs, Dirac's positive energy relativistic wave equation, Weyl-Schwinger unitary basis for 
matrix algebra and Alladi Ramakrishnan's matrix decomposition theorem, finite-dimensional 
Wigner function, finite-dimensional canonical transformations, magnetic Bloch functions, 
finite-dimensional quantum mechanics, and the relation between GCAs and quantum groups.   
\end{abstract}

\vfill 
\noindent
{\sf To appear in}\,: ``{\em The Legacy of Alladi Ramakrishnan in the Mathematical Sciences}'' 
(K. Alladi, J. Klauder, C. R. Rao, Editors) Springer, 2010. 

\newpage 
\section{Introduction} 
\renewcommand\theequation{\thesection.\arabic{equation}}
\setcounter{equation}{0}
Extensive studies on Clifford algebra, its generalizations, and their physical applications 
were made for about a decade starting 1967, under the name of $L$-Matrix Theory, by Alladi 
Ramakrishnan and his collaborators at The Institute of Mathematical Sciences (MATSCIENCE) including 
me, his Ph.D student during 1971-76.  When I joined MATSCIENCE in August 1971, as a student, 
the book \cite{AR}, containing all the results of their papers on the subject, up to then, was 
getting ready to be released; I had participated in the final stage of proof reading of the book.  
Chandrasekaran had just completed his Ph.D. thesis on the topic \cite{PSC}.  Subsequently, I 
started my thesis work on the same topic under the guidance of Alladi Ramakrishnan.   I had 
also the guidance of Ranganathan,  Santhanam and Vasudevan, senior faculty members of the 
institute, who had also started their scientific careers under the guidance of Alladi Ramakrishnan 
and had contributed largely to the development of $L$-matrix theory.  In my Ph.D. thesis \cite{J1} 
I had studied certain group theoretical aspects of generalized Clifford algebras (GCAs) and 
their physical applications.  After my Ph.D. work also, I have applied the elements of GCAs in 
studies of certain problems in quantum mechanics, and quantum groups.  Here, I would like to 
outline some aspects of GCAs and their applications essentially based on my work.   

A generalized Clifford algebra (GCA) can be presented, in general, as an algebra having a basis 
with generators $\{e_j | j = 1,2,\ldots,n\}$ satisfying the relations 
\bea 
e_j e_k & = & \o_{jk} e_k e_j, \quad 
\o_{jk} e_l\,=\,e_l \o_{jk}, \quad 
\o_{jk}\o_{lm}\,=\,\o_{lm}\o_{jk}, \nn \\
e_j^{N_j} & = & 1, \quad 
\o_{jk}^{N_j}\,=\,\o_{jk}^{N_k}\,=\,1, \qquad \forall\ \ j,k,l,m\,=\,1,2,\ldots,n.  
\lb{gca} 
\eea 
In any irreducible matrix representation, relevant for physical applications, one will have   
\bea 
\o_{jk} & = & \o_{kj}^{-1}\,=\,e^{2\pi i \nu_{jk}/N_{jk}}, \quad 
N_{jk}\,=\,\mbox{g.c.d}\,(N_j , N_k),  \nn \\ 
           &     &  \quad \quad j,k = 1,2,\ldots,n 
\lb{gca2}
\eea  
where $\nu_{jk}$s are integers.  Consequently, one can write 
\bea
\o_{jk} & = & e^{2\pi i t_{jk}/{\h{N}}}, \quad 
t_{kj}\,=\,-t_{jk}, \ \ \h{N}\,=\,\  \mbox{l.c.m}\,[N_{jk}], \nn \\ 
          &     &  \quad j,k = 1,2,\ldots,n .  
\lb{gca3} 
\eea 
Thus, any GCA can be characterized by an integer $\h{N}$ and an antisymmetric integer matrix 
\bea
T & = & 
\lrb\ba{ccccc}
        0 & t_{12} & t_{13} & \ldots & t_{1n} \\
-t_{12} &    0     & t_{23} & \ldots & t_{2n} \\
-t_{13} & -t_{23} &   0     & \ldots & t_{3n} \\ 
\vdots & \vdots   & \vdots & \ddots & \vdots \\ 
-t_{1n} & -t_{2n} & -t_{3n} & \ldots &  0 \\
\ea\rrb. 
\lb{T} 
\eea
In the following we shall study the representation theory of GCAs and physical applications of 
some special cases of these algebras.    

\section{Projective representations of finite abelian groups and GCAs} 
\renewcommand\theequation{\thesection.\arabic{equation}}
\setcounter{equation}{0}
GCAs arise in the study of projective, or ray, representations of finite abelian groups.   
Let us consider the finite abelian group 
$G \cong \mathbb{Z}_{N_1}\ox \mathbb{Z}_{N_2}\ox \ldots \ox \mathbb{Z}_{N_n}$ with 
$\{c_1^{m_1}c_2^{m_2}\ldots c_n^{m_n}\}$ as its generic element where the generators $\{c_j\}$ 
satisfy the relations 
\bea
c_j c_k & = & c_k c_j,  \quad c_j^{N_j} = 1, \qquad j\,=\,1,2,\ldots,n. 
\eea 
A projective representation $D(G)$ of a group $G$ is defined as 
\bea
D(g_j)D(g_k) & = & \vf (g_j , g_k) D(g_jg_k), \ \ \vf (g_j , g_k) \in \mathbb{C}, 
\qquad \forall \  g_j , g_k \in G, \nn \\ 
   &     &    
\lb{projrep}
\eea 
where the given factor set $\{ \vf (g_j , g_k) \}$ is such that 
\bea
\vf (g_j , g_k)\vf (g_jg_k , g_l) & = \vf (g_j , g_kg_l) \vf (g_k , g_l),  
\qquad \forall \  g_j , g_k , g_l \in G, 
\lb{vf}
\eea  
and 
\bea
\vf (E , g_j) & = & \vf (g_j , E)\,=\,1, \qquad \forall \  g_j \in G, 
\lb{idvf} 
\eea 
with $E$ as the identity element of $G$.   For an abelian group, equation (\ref{projrep}) implies 
\bea
D(g_j) D(g_k) & = & \vf (g_j , g_k) D(g_jg_k)\,=\,\vf (g_j , g_k) D(g_kg_j) \nn \\ 
                       & = & \fr{\vf (g_j , g_k)}{\vf (g_k , g_j)} D(g_k)D(g_j), 
\qquad \forall \  g_j , g_k \in G, 
\lb{prabelian}
\eea
or, 
\bea
D(g_j) D(g_k) & = & \O_\vf (g_j , g_k) D(g_k) D(g_j), 
\lb{comrel}
\eea 
with 
\bea 
\O_\vf (g_j , g_k)\,=\,\fr{\vf (g_j, g _k)}{\vf (g_k, g _j)}, \qquad \forall \  g_j,g_k \in G.  
\lb{O} 
\eea 
Using (\ref{projrep}) it is easy to see that we can write 
\bea
D \lrb \Pi_{j=1}^n c_j^{m_j} \rrb & = & 
\phi \lrb \Pi_{j=1}^n c_j^{m_j} \rrb \lcb \Pi_{j=1}^n D(c_j)^{m_j} \rcb,  
\lb{rep}
\eea 
with 
\bea 
\phi \lrb \Pi_{j=1}^n c_j^{m_j} \rrb & = & 
  \Pi_{j=1}^n \Pi_{ p_j=1}^{m_j} \vf \lrb c_j,\Pi_{l=0}^{n-j} 
    c_j^{N_j-p_j}c_{j+l}^{m_{j+l}} \rrb^{-1}. 
\lb{phi} 
\eea  
From this it follows that 
\bea 
D \lrb c_j^{N_j} \rrb & = & \phi \lrb c_j^{N_j} \rrb D(c_j)^{N_j}\,=\,I, 
\qquad \forall\ j = 1,2,\ldots,n., 
\eea 
where 
\bea 
\phi \lrb c_j^{N_j} \rrb & = & \Pi_{ p_j=1}^{N_j} \vf  \lrb c_j , c_j^{N_j - p_j} \rrb^{-1}.  
\eea 
Let us now define 
\bea
e_j & = & \phi \lrb c_j^{N_j} \rrb^{1/N_j} D(c_j), \qquad \forall\ j = 1,2,\ldots,n. 
\eea 
Then, it is found that the required representations satisfying (\ref{projrep}-\ref{O}), for 
the given factor set, are immediately obtained from (\ref{rep}-\ref{phi}) once the ordinary 
representations of $\{ e_j | j = 1,2,\ldots,n \}$ are found such that 
\bea
e_j e_k & = & \o^{(\vf)}_{jk} e_k e_j, \quad 
\o^{(\vf)}_{jk} e_l\,=\,e_l \o^{(\vf)}_{jk}, \quad 
\o^{(\vf)}_{jk}\o^{(\vf)}_{lm}\,=\,\o^{(\vf)}_{lm}\o^{(\vf)}_{jk}, \nn \\ 
e_j^{N_j} & = & 1,  \quad 
\lrb \o^{(\vf)}_{jk} \rrb^{N_j}\,=\,\lrb \o^{(\vf)}_{jk} \rrb^{N_k}\,=\,1, 
\quad \mbox{with}\ \ \o^{(\vf)}_{jk}\,=\,\O_\vf (c_j , c_k), \nn \\ 
   &   &  \qquad \qquad \forall\ j,k,l,m = 1,2,\ldots,n.    
\lb{ejek}
\eea
Comparing (\ref{ejek}) with (\ref{gca}) it is clear that the problem of finding the projective 
representations of any finite abelian group for any given factor set reduces to the problem of 
finding the ordinary representations of a generalized Clifford algebra defined by (\ref{gca}).  

\section{Representations of GCAs}  
\renewcommand\theequation{\thesection.\arabic{equation}}
\setcounter{equation}{0}
Let us now consider a GCA associated with a specific antisymmetric integer matrix $T$ as in 
(\ref{T}) and an integer $\h{N}$.  The  $T$-matrix can be related to its skew-normal form 
${\cal T}$ by a transformation as follows\,:  
\bea
\T & = & 
\lrb\ba{cc} 
0     & t_1 \\
-t_1 & 0   \\
\ea\rrb \oplus \cdots \oplus  
\lrb\ba{cc}
0     & t_s \\ 
-t_s & 0    \\ 
\ea\rrb \oplus 
{\mbox O}_{n-2s}, \nn \\
T & = & U {\cal T} \tilde{U}\,(\pm\mbox{mod}.\h{N}),  
\lb{skewT}
\eea 
where ${\mbox O}_{n-2s}$ is an $(n-2s) \times (n-2s)$ null matrix, $U = [u_{jk}]$ is a 
unimodular integer matrix with $|u_{jk}| \leq \h{N}$, and $\tilde{U}$ is the transpose of $U$.   
For any given antisymmetric integer matrix $T$ it is possible to get the skew normal form $\T$ 
and the corresponding $U$-matrix explicitly by a systematic procedure (see, {\em e.g.}, 
\cite{N}).  Now, let $\{ \e_j | j = 1,2,\ldots,n\}$ be a set of elements satisfying the 
commutation relations 
\bea
\e_{2j-1} \e_{2j} & = & e^{2\pi i t_j/\h{N}} \e_{2j} \e_{2j-1}, 
\qquad j = 1,2,\ldots,s, \nn \\ 
\e_k \e_l & = & \e_l \e_k \ \  \mbox{otherwise}.    
\lb{Trel}
\eea 
It is clear that this set of relations generate a GCA corresponding to $\T$ as its $T$-matrix.  
It is straightforward to verify that if we construct $\{e_j | j = 1,2,\ldots,n \}$ from 
$\{ \e_j | j = 1,2,\ldots,n \}$ through a product transformation \cite{J1,RJ} 
\bea
e_j & = & \mu_j \e_1^{u_{j1}}\e_2^{u_{j2}}\ldots\e_n^{u_{jn}}, 
\qquad \forall\ j = 1,2,\ldots,n,  
\lb{prodtrans}
\eea  
where $\lsb u_{jk} \rsb = U$ and $\{ \mu_j | j = 1,2,\ldots,n \}$ are complex numbers, then, 
in view of (\ref{skewT}), 
\bea 
e_j e_k & = & e^{2\pi i t_{jk}/\h{N}} e_k e_j, \qquad \forall\ j,k = 1,2,\ldots,n., 
\eea 
as required in (\ref{gca})-(\ref{gca3}); the complex numbers $\{ \mu_j \}$ are normalization 
factors which are to be chosen such that 
\bea 
e_j^{N_j} & = & 1, \qquad \forall\ j = 1,2,\ldots,n. 
\eea 

Now, let the matrix representations of $\{ \e_j | j = 1,2,\ldots,2s \}$ be given by 
\bea 
\e_1 & = & I \ox I \ox I \ox \cdots \ox I \ox A_1, \nn \\ 
\e_2 & = & I \ox I \ox I \ox \cdots \ox I \ox B_1, \nn \\
\e_3 & = & I \ox I \ox I \ox \cdots \ox A_2 \ox I , \nn \\ 
\e_4 & = & I \ox I \ox I \ox \cdots \ox B_2 \ox I , \nn \\
   & \vdots &   \nn \\
\e_{2s-1} & = & A_s \ox I \ox I \ox \cdots \ox I \ox I , \nn \\ 
\e_{2s} & = & B_s \ox I \ox I \ox \cdots \ox I \ox I, 
\eea 
where 
\bea
A_j B_j & = & \o_j^{\t_j} B_j A_j, \nn \\ 
&   &  \ \ \mbox{with}\ \ \o_j\,=\,e^{2\pi i/{\mbox{N}_j}},  
           \quad {\mbox{N}}_j\,=\,\h{N}/(\mbox{g.c.d.}(t_j , \h{N})), \nn \\ 
 &   & \ \ \t_j\,=\,t_j/(\mbox{g.c.d}(t_j , \h{N})), \qquad   j = 1,2,\ldots,s, 
\lb{AjBjs} 
\eea 
and $I$s are identity matrices of appropriate dimensions.  Since 
$\{ \e_k | k = 2s+1,2s+2,\ldots,n \}$ commute among themselves and also with all 
other $\{ e_j | j = 1,2,\ldots,2s \}$ they are represented by unimodular complex numbers which 
can be absorbed in the normalization factors $\{ \mu_j \}$ in (\ref{prodtrans}).    This shows 
that if the matrix representations of all $A$s and $B$s satisfying (\ref{AjBjs}) are known, then 
the problem of representation of the given GCA is solved.  Explicitly, one has, apart from 
multiplicative normalizing phase factors,  
\bea
e_j & \sim & A_s^{u_{j(2s-1)}} B_s^{u_{j(2s)}} \ox A_{s-1}^{u_{j(2s-3)}} B_{s-1}^{u_{j(2s-2)}} 
                      \ox \ldots \ox A_1^{u_{j1}} B_1^{u_{j2}}, \nn \\ 
      &         & \qquad \qquad  \forall\ j = 1,2,\ldots,n.  
\lb{explrep}
\eea 
Note that $\o_j^{\t_j}$s in (\ref{AjBjs}) are primitive roots of unity. Thus the representation 
theory of any GCA depends essentially on the central relation 
\bea
AB & = & \o BA, 
\lb{weyl} 
\eea  
where $\o$ is a nontrivial primitive root of unity.  If $\o$ is a primitive $N$th root of 
unity then the normalization relations for $A$ and $B$ can be 
\bea 
A^{jN} & = & I, \quad B^{kN}\,=\, I, \ \ \mbox{where}\ j,k = 1,2,\ldots.  
\lb{norm} 
\eea 
The central relation (\ref{weyl}) determines the representation of $A$ and $B$ uniquely up to 
multiplicative phase factors and the normalization relation (\ref{norm}) fixes these phase 
factors.  For more details on projective representions of finite abelian groups and their relation 
to GCAs, and other different approaches to GCAs, see \cite{MN}-\cite{J2}.  

\section{The Clifford algebra}  
\renewcommand\theequation{\thesection.\arabic{equation}}
\setcounter{equation}{0} 
Hamilton's quaternion, generalizing the complex number, is given by 
\bea 
\mq & = & q_0 1 + q_1\mi + q_2\mj + q_3\mk, 
\eea  
where $\{q_0,q_1,q_2,q_3\}$ are real numbers, $1$ is the identity unit, and $\{\mi,\mj,\mk\}$ 
are imaginary units such that  
\bea
\mi\mj & = & -\mj\mi, \quad \mj\mk\,=\,-\mk\mj, \quad \mk\mi\,=\,-\mi\mk, \nn \\ 
\mi^2 & = & \mj^2\,=\,\mk^2\,=\,-1, 
\lb{qcr} 
\eea 
and 
\bea
\mi\mj & = & \mk, \quad \mj\mk\,=\,\mi, \quad \mk\mi\,=\,\mj.  
\lb{qnr} 
\eea  
It should be noted that the relations in (\ref{qnr}) are not independent of the commutation 
and normalization relations  (\ref{qcr}); to see this, observe that $\mi\mj\mk$ commutes with 
each one of the imaginary units $\{\mi,\mj,\mk\}$ and hence $\mi\mj\mk \sim 1$.  The `geometric 
algebra' of Clifford \cite{C} has the generating relations  
\bea
\iota_j\iota_k & = & -\iota_k\iota_j, \qquad \mbox{for}\ j \neq k \nn \\
\iota_j^2 & = & -1, \qquad \forall\ j,k = 1,2,\ldots,n,    
\lb{GA} 
\eea 
obtained by generalizing (\ref{qcr}).  This is what has become the Clifford algebra defined by 
the generating relations 
\bea
e_j e_k & = & -e_k e_j, \qquad \mbox{for}\ j \neq k, \nn \\
e_j^2 & = & 1, \qquad \forall\ j,k = 1,2,\ldots,n,    
\lb{CA} 
\eea 
which differ from (\ref{GA}) only in the normalization conditions, and evolved into the GCA 
(\ref{gca}).  Thus, the Clifford algebra (\ref{CA}) corresponds to (\ref{gca}) with the choice 
\bea 
\o_{jk} & = & -1, \quad N_j\,=\,2, \qquad \forall\ j,k = 1,2,\ldots,n, 
\eea 
associated with the $T$-matrix 
\bea 
T & = & 
\lrb\ba{ccccc}
        0 & 1 & 1 & \ldots & 1 \\ 
-1 &    0     & 1 & \ldots & 1 \\
-1 & -1 &   0     & \ldots & 1 \\ 
\vdots & \vdots   & \vdots & \ddots & \vdots \\ 
-1 & -1 & -1 & \ldots &  0 \\
\ea\rrb.   
\lb{CAT} 
\eea
and 
\bea 
\h{N} = 2. 
\eea   
The corresponding skew normal form is  
\bea
\T & = & 
\underbrace{ 
\lrb\ba{cc} 
0     & 1 \\
-1 & 0   \\
\ea\rrb \oplus 
\lrb\ba{cc}
0 & 1 \\
-1 & 0 \\
\ea\rrb \oplus 
\cdots \oplus  
\lrb\ba{cc}
0     & 1 \\ 
-1 & 0    \\ 
\ea\rrb}_{m\ \mbox{times}},  
\lb{catau2m}
\eea
when $n = 2m$.  When $n = 2m+1$, 
\bea
\T & = & 
\underbrace
{\lrb\ba{cc} 
0     & 1 \\
-1 & 0   \\
\ea\rrb \oplus 
\lrb\ba{cc}
0 & 1 \\
-1 & 0 \\
\ea\rrb \oplus 
\cdots \oplus  
\lrb\ba{cc}
0     & 1 \\ 
-1 & 0    \\ 
\ea\rrb}_{m\ \mbox{times}}
\oplus\ 0.   
\lb{catau2m+1}
\eea
In this case the $U$-matrices are 
\bea
U & = & \lrb\ba{ccccccccc} 
0 & 0 & 0 & 0 & \ldots & 0 & 0 & 1 & 0 \\
0 & 0 & 0 & 0 & \ldots & 0 & 0 & 0 & 1 \\
0 & 0 & 0 & 0 & \ldots & 1 & 0 & -1 & 1 \\
0 & 0 & 0 & 0 & \ldots & 0 & 1 & -1 & 1 \\
\vdots & \vdots & \vdots & \vdots & \ddots & \vdots & \vdots & \vdots & \vdots \\ 
1 & 0 & -1 & 1 & \ldots & -1 & 1 & -1 & 1 \\
0 & 1 & -1 & 1 & \ldots & -1 & 1 & -1 & 1 \\
\ea\rrb, \nn \\ 
   &   &  \qquad \qquad \mbox{for}\ n = 2m, 
\lb{2mU} 
\eea  
and 
\bea 
U & = & \lrb\ba{cccccccccc} 
0 & 0 & 0 & 0 & \ldots & 0 & 0 & 1 & 0 & 0 \\
0 & 0 & 0 & 0 & \ldots & 0 & 0 & 0 & 1 & 0 \\
0 & 0 & 0 & 0 & \ldots & 1 & 0 & -1 & 1 & 0 \\
0 & 0 & 0 & 0 & \ldots & 0 & 1 & -1 & 1 & 0 \\
\vdots & \vdots & \vdots & \vdots & \ddots & \vdots & \vdots & \vdots & \vdots & \vdots \\ 
1 & 0 & -1 & 1 & \ldots & -1 & 1 & -1 & 1 & 0 \\
0 & 1 & -1 & 1 & \ldots & -1 & 1 & -1 & 1 & 0 \\ 
-1 & 1 & -1 & 1 & \ldots & -1 & 1 & -1 & 1 & 1 \\ 
\ea\rrb, \nn \\ 
   &   &  \qquad \qquad \mbox{for}\ n = 2m+1, 
\lb{2m+1U} 
\eea
such that 
\bea 
T & = & U{\cal T} \tilde{U}\,(\pm\mbox{mod}.2). 
\eea 
Now, equation (\ref{Trel}) becomes in this case, for both $n = 2m$ and $n = 2m+1$,   
\bea
\e_{2j-1} \e_{2j} & = & - \e_{2j} \e_{2j-1}, \qquad j = 1,2,\ldots,m, \nn \\ 
\e_k \e_l & = & \e_l \e_k,  \qquad \mbox{otherwise}, 
\lb{caTrel}
\eea 
with the matrix representations 
\bea 
\e_1 & = & I \ox I \ox I \ox \cdots \ox I \ox A_1, \nn \\ 
\e_2 & = & I \ox I \ox I \ox \cdots \ox I \ox B_1, \nn \\
\e_3 & = & I \ox I \ox I \ox \cdots \ox A_2 \ox I , \nn \\ 
\e_4 & = & I \ox I \ox I \ox \cdots \ox B_2 \ox I, \nn \\
   & \vdots &   \nn \\
\e_{2m-1} & = & A_m \ox I \ox I \ox \cdots \ox I \ox I, \nn \\ 
\e_{2m} & = & B_m \ox I \ox I \ox \cdots \ox I \ox I, \nn \\
\eea 
where 
\bea 
A_j & = & \s_1\,=\,\lrb\ba{cc} 0 & 1 \\ 1 & 0 \\ \ea\rrb, \ \ 
B_j\,=\,\s_3\,=\,\lrb\ba{cr} 1 & 0 \\ 0 & -1 \\ \ea\rrb, \nn \\ 
       &     &  \qquad \qquad \forall\ j = 1,2,\ldots,m. 
\eea 
In the case of $n = 2m+1$, since $\e_{2m+1}$ commutes with all other $\e_j$s it can be just 
taken to be $1$.   The matrices $\s_1$ and $\s_3$ are the well known first and the third Pauli 
matrices, respectively, and the second Pauli matrix is given by 
\bea 
\s_2 & = & i\s_1\s_3\,=\,\lrb\ba{cr} 0 & -i \\ i & 0 \\ \ea\rrb. 
\eea 
Then, in view of (\ref{explrep}) and (\ref{2mU},\ref{2m+1U}), the required representations of 
(\ref{CA}) are given in terms of the Pauli matrices 
by 
\bea 
e_1 & = & \s_1 \ox I \ox \cdots \ox I \ox I, \nn \\ 
e_2 & = & \s_3 \ox I \ox \cdots \ox I \ox I, \nn \\ 
e_3 & = & \s_2 \ox \s_1 \ox I \ox \cdots \ox I \ox I, \nn \\ 
e_4 & = & \s_2 \ox \s_3 \ox I \ox \cdots \ox I \ox I \nn \\ 
   & \vdots &  \nn \\ 
e_{2m-1} & = & \s_2 \ox \s_2 \ox \cdots \ox \s_2 \ox \s_1, \nn \\   
e_{2m} & = & \s_2 \ox \s_2 \ox \cdots \ox \s_2 \ox \s_3, \nn \\    
e_{2m+1} & = & \s_2 \ox \s_2 \ox \cdots \s_2 \ox \s_2.  
\lb{careps} 
\eea 
Note that this representation is Hermitian and unitary.  One can show that this is an 
irreducible representation.  Also it should be noted that the above representation matrices 
are defined only upto multiplication by $\pm 1$ since $e_j^2 = 1$ for all $j$.  

Let us now write down the generators of the first four Clifford algebras\,: 
\bea  
C^{(2)} & : & e_1^{(2)}\,=\,\s_1, \ \ e_2^{(2)}\,=\,\s_3, \nn \\ 
C^{(3)} & : & e_1^{(3)}\,=\,\s_1, \ \ e_2^{(3)}\,=\,\s_3, \ \ e_3^{(3)}\,=\,\s_2, \nn \\ 
C^{(4)} & : & e_1^{(4)}\,=\,\s_1 \ox I, \ \ e_2^{(4)}\,=\,\s_3 \ox I, \nn \\  
             &   & e_3^{(4)}\,=\,\s_2 \ox \s_1, \ \ e_4^{(4)}\,=\,\s_2 \ox \s_3, \nn \\ 
C^{(5)} & : & e_1^{(5)}\,=\,\s_1 \ox I, \ \ e_2^{(5)}\,=\,\s_3 \ox I, \nn \\ 
             &   & e_3^{(5)}\,=\,\s_2 \ox \s_1, \ \ e_4^{(5)}\,=\,\s_2 \ox \s_3, \ \ 
                   e_5^{(5)}\,=\,\s_2 \ox \s_2,  
\lb{fewcareps}
\eea 
where the superscript indicates the number of generators in the corresponding algebra.  The 
dimension of the irreducible representation of the Clifford algebra with $2m$, or $2m+1$, 
generators is $2^m$.  One can show that for the algebra with an even number of generators 
there is only one unique irreducible representation up to equivalence.  In the case of the 
algebra with an odd number of generators there are two inequivalent irreducible representations 
where the other representation is given by multiplying all the matrices of the first 
representation by $-1$.   These statements form Pauli's theorem on Clifford algebra.  

An obvious irreducible representation of the identity and the three imaginary units of Hamilton's 
quaternion algebra (\ref{qcr},\ref{qnr}) is given by  
\bea
1 & = & I, \quad \mi\,=\,-i\s_1, \quad \mj\,=\,-i\s_3, \quad \mk\,=\,i\s_2.
\eea 
From the above it is clear that, as Clifford remarked \cite{C}, the geometric algebra, or the 
Clifford algebra, is a compound of quaternion algebras the units of which are commuting with one 
another.  Actually, the equations (\ref{Trel}) and (\ref{prodtrans}) correspond precisely to 
Clifford's original construction of geometric algebra starting with commuting quaternion algebras; 
matrix representations and realization of commuting quaternion algebras in terms of direct 
products did not exist at that time.  Later, obviously unaware of Clifford's work, Dirac 
\cite{D1} used the same procedure to construct his four matrices $\{ \a_x, \a_y, \a_z, \b \}$, 
building blocks of his relativistic theory of electron and other spin-$1/2$  particles, starting 
with the three Pauli matrices $\{\s_1,\s_2,\s_3\}$.  The Dirac matrices are given by  
\bea
\a_x & = & \s_1 \ox \s_1, \ \ \a_y\,=\,\s_1 \ox \s_2, \ \ \a_z\,=\,\s_1 
           \ox \s_3, \ \ \b\,=\,\s_3 \ox I, 
\eea 
which can be shown to be equivalent to the representation of $C^{(4)}$ given above; as already 
mentioned, $C^{(4)}$ has only one inequivalent irreducible representation.   Clifford algebra 
is basic to the theory of spinors, theory of fermion fields, Onsager's solution of the two 
dimensional Ising model, etc.   For detailed accounts of Clifford algebra and its various physical 
applications see, {\em e.g.}, \cite{H}-\cite{DL}.   

\section{Alladi Ramakrishnan's $L$-matrix theory and $\sigma$-operation}  
\renewcommand\theequation{\thesection.\arabic{equation}}
\setcounter{equation}{0}
Representation theory of Clifford algebra has been expressed by Alladi Ramakrishnan \cite{AR} 
in a very nice framework called the $L$-matrix theory.  Let 
\bea
L^{(2m+1)}(\u{\l}) & = & \sum_{j=1}^{2m+1} \l_j e_j^{(2m+1)}, 
\eea 
called an $L$-matrix, be associated with a $(2m+1)$-dimensional vector 
$\u{\l} = (\l_1,\l_2,\ldots,\l_{2m+1})$.  It follows that 
\bea
\lrb L^{(2m+1)}(\u{\l}) \rrb^2 & = & \lrb \sum_{j=1}^{2m+1} \l_j^2 \rrb I\,=\,||\u{\l}||^2 I, 
\eea 
where $I$ is the $2^m \times 2^m$ identity matrix.  Thus, $L^2$ represents the square of the 
norm, or the length, of the vector $\u{\l}$.  In other words, $L$ is a square root of 
$\sum \l_j^2$ linear in $\{ \l_j \}$.  

From (\ref{fewcareps}) observe that 
\bea
e_1^{(5)} & = & e_1^{(3)} \ox I, \quad 
e_2^{(5)}\,=\,e_2^{(3)} \ox I, \nn \\
e_3^{(5)} & = & e_3^{(3)} \ox e_1^{(3)}, \quad 
e_4^{(5)}\,=\,e_3^{(3)} \ox e_2^{(3)}, \quad 
e_5^{(5)}\,=\,e_3^{(3)} \ox e_3^{(3)}.  
\eea 
Thus, one can write 
\bea
L^{(5)}(\u{\l}) & = & 
  \sum_{j=1}^5 \l_j e_j^{(5)} \nn \\ 
    & = & e_1^{(3)} \ox \l_1I + e_2^{(3)} \ox \l_2I 
        + e_3^{(3)} \ox \lrb \l_3 e_1^{(3)}+ \l_4 e_2^{(3)}+ \l_5 e_3^{(3)} \rrb,  \nn \\ 
                        &     &    
\eea 
{\em i.e.}, $L^{(5)}$ can be obtained from $L^{(3)}$ by replacing $\l_1$, $\l_2$, and $\l_3$ by 
$\l_1I$, $\l_2I$, and $L^{(3)}(\l_3,\l_4,\l_5)$, respectively.   From (\ref{careps}) it is 
straightforward to see that this procedure generalizes\,: an $L^{(2m+3)}$ can be obtained from 
an $L^{(2m+1)}$ by replacing $(\l_1,\l_2,\ldots,\l_{2m})$, respectively, by 
$(\l_1I,\l_2I,\ldots,\l_{2m}I)$, and $\l_{2m+1}$ by $L^{(3)}(\l_{2m+1},\l_{2m+2},\l_{2m+3})$.  
This procedure is called $\sigma$-operation by Alladi Ramakrishnan.  It can be shown that the 
induced representation technique of group theory takes this form in the context of Clifford 
algebra \cite{ARIVVR}.  Actually, in this prcedure any one of the parameters of $L^{(2m+1)}$ 
can be replaced by an $L^{(3)}$ and the remaining parameters $\{\l_j\}$ can be replaced, 
respectively, by $\{\l_jI\}$ with suitable relabelling.   As we shall see below this 
$\sigma$-operation generalizes to the case of GCAs with ordered $\o$-commutation relations.  

Another interesting result of Alladi Ramakrishnan is on the  diagonalization of an $L$-matrix.  
An $L^{(2m+1)}$-matrix of dimension $2^m$ obeys 
\bea
\lrb L^{(2m+1)} \rrb^2 & = & \sum_{j=1}^{2m+1} \l_j^2 I\,=\,\L^2 I, 
\eea 
and hence has $(\L, -\L)$ as its eigenvalues each being $2^{m-1}$-fold degenerate.  In general, 
let us call the matrix $e_2^{(2m+1)}$, or $e_2^{(2m)}$,  as $\b$\,:
\bea
\b & = & \lrb \ba{rr} I & 0 \\ 0 & -I \\ \ea \rrb, 
\eea 
where $I$ is the $2^{m-1}$-dimensional identity matrix.  Thus, the diagonal form of $L$ is $\L\b$.  
Then, from the relation 
\bea
L(L + \L\b) & = & \L^2I + L\L\b\,=\,(L + \L\b)\L\b, 
\eea
it follows that $(L + \L\b)$ is the matrix diagonalizing $L$ and the columns of $(L + \L\b)$ 
are the eigenvectors of $L$. Note that an $L^{(2m)}$-matrix, of dimension $2^m$, can be 
treated as an $L^{(2m+1)}$-matrix with one of the $\l$s as zero.  

Let us now take a Hermitian $L(\u{\l})$-matrix where all the $\l$-parameters are real.  Since 
$\b = e_2$ anticommutes with all the other $e_j$s we get 
\bea
(L + \L\b)^2 & = & 2\L^2I + \L(L\b + \b L) = 2\L(\L + \l_2)I. 
\eea 
Hence 
\bea
U & = & \fr{L + \L\b}{\sqrt{2\L(\L+\l_2)}}
\eea 
is Hermitian and unitary ($U = U^\dagger = U^{-1}$) and is such that 
\bea
U^{-1}LU & = & \L\b. 
\eea 
Thus, the columns of $U$ are normalized eigenvectors of the Hermitian $L$.  This result has 
been applied \cite{AR} to solve in a very simple manner Dirac's relativistic wave equation 
\cite{D1}, 
\bea
i \hbar \fr{\p\psi(\v{r},t)}{\p t} & = & \lsb -i \hbar c \lrb \a_x\fr{\p}{\p x} 
    + \a_y\fr{\p}{\p y} + \a_z\fr{\p}{\p z} \rrb + mc^2\b \rsb \psi(\v{r},t), \nn \\ 
   &    &  
\lb{oDeq} 
\eea  
where $\psi(\v{r},t)$ is the $4$-component spinor associated with the free spin-$1/2$ particle.   

\section{Dirac's positive-energy relativistic wave equation}  
\renewcommand\theequation{\thesection.\arabic{equation}}
\setcounter{equation}{0}
Students of Alladi Ramakrishnan got excellent training as professional scientists.  He emphasized 
that the students should master any topic of research by studying the works of the leaders in 
the field and should communicate with their peers whenever necessary.  In this connection, I 
would like to recall proudly an incident.  

Following a suggestion of Santhanam, my fellow junior student Dutt and I started studying a 
paper of Dirac \cite{D2} 
in which he had proposed a positive-energy relativistic wave equation\,: 
\bea
    &     &  i \hbar \fr{\p \lsb \h{q} \psi(\v{r},t;q_1,q_2) \rsb}{\p t} \nn \\ 
    &     &   \qquad =\,\lsb -i \hbar c \lrb \a_x'\fr{\p}{\p x} + \a_y'\fr{\p}{\p y}  
             + \a_z'\fr{\p}{\p z} \rrb 
             + mc^2\b' \rsb \lsb \h{q} \psi(\v{r},t;q_1,q_2) \rsb, \nn \\ 
    &     & 
\lb{nDeq}
\eea 
$\lsb\h{q}\psi\rsb$ being a $4$-component column matrix with elements 
$(\h{q}_1\psi, \h{q}_2\psi, \h{q}_3\psi, \h{q}_4\psi)$  where 
\bea
\lsb \h{q}_j , \h{q}_k \rsb & = & -\beta'_{jk}, \qquad j,k\,=\,1,2,3,4, 
\lb{Diracqq} 
\eea 
and 
\bea
\b' & = & \s_2 \ox I, \quad 
\a_x'\,=\,-\s_1 \ox \s_3, \quad 
\a_y'\,=\,\s_1 \ox \s_1, \quad 
\a_z'\,=\,\s_3 \ox I.  \nn \\ 
   &     &   
\eea 
Unlike the standard relativistic wave equation for the electron (\ref{oDeq}) which has both 
positive and negative (antiparticle) energy solutions, the new Dirac equation (\ref{nDeq}) has 
only positive energy solutions.  Further, more interestingly, this positive-energy particle 
would not interact with an electromagnetic field.  Around November 1974, Dutt and I stumbled 
upon an equation which had only negative-energy solutions.   Our negative-energy relativistic 
wave equation was exactly the same as Dirac's positive-energy equation (\ref{nDeq}) except only 
for a slight change in the commutation relations of the internal variables 
$(\h{q}_1,\h{q}_2,\h{q}_3,\h{q}_4)$ in the equation\,; instead of (\ref{Diracqq}), we took 
\bea 
\lsb \h{q}_j , \h{q}_k \rsb & = & \beta'_{jk}, \quad j,k\,=\,1,2,3,4. 
\eea 
When I told Alladi Ramakrishnan about this he told us that we could not meddle with Dirac's 
work and keep quiet.  He suggested that I  should write to Dirac and get his opinion on our 
work.  I wrote to Dirac who was in The Florida State University at that time.  I received a 
letter from him within a month!  His reply was\,:  ``{\sl Dear Jagannathan, The equation you 
propose would correctly describe a particle with only negative-energy states.  It would be the 
correct counterpart of the positive-energy equation, but of course it would not have any physical 
application. Yours sincerely,  P. A. M. Dirac.}''  Immediately, Alladi Ramakrishnan forwarded 
our paper for rapid publication \cite{JD}.  

So far, no one has found any application for Dirac's positive-energy equation.  Attempts to 
modify it so that these positive-energy particles could interact with electromagnetic field 
have not succeeded.  May be, these positive-energy Dirac particles and their negative-energy 
antiparticles constitute the dark matter of our universe.  

\section{GCAs with ordered $\omega$-commutation relations}  
\renewcommand\theequation{\thesection.\arabic{equation}}
\setcounter{equation}{0}
We shall now consider a GCA (\ref{gca}) with ordered $\o$-commutation relations, {\em i.e.}, 
\bea
e_je_k & = & \o e_ke_j, \quad \o\,=\,e^{2\pi i/N}, \quad \forall\ j < k, \nn \\ 
e_j^N & = & 1, \qquad j,k\,=\,1,2,\ldots,n.   
\lb{ordgca} 
\eea  
The associated $T$-matrix has elements 
\bea 
t_{jk} & = & \lcb \ba{rl} 1, & \mbox{for}\ j < k, \\ 
                          0, & \mbox{for}\ j\,=\,k, \\ 
                         -1, & \mbox{for}\ j > k, \\ 
             \ea \right. 
\eea 
and $\h{N} = N$.  This is exactly same as for the Clifford algebra except for the value of 
$\h{N}$.  So, the treatment  of representation theory of this GCA is along the same lines as 
for the Clifford algebra: $T$ matrix is the same as in (\ref{CAT}) for any $n$ and $\cal{T}$ 
and $U$ matrices are the same as in (\ref{catau2m}) and (\ref{2mU}) for $n = 2m$ and 
(\ref{catau2m+1}) and (\ref{2m+1U}) for $n = 2m+1$, respectively.  The only difference is that 
in the case of Clifford algebra $A_j^{-1} = A_j$ and $B_j^{-1} = B_j$ for any $j$ where as now 
$A_j^{-1} = A_j^{N-1}$ and $B_j^{-1} = B_j^{N-1}$ for any $j$.  Thus, in view of (\ref{explrep}) 
and (\ref{2mU},\ref{2m+1U}), the required representations of (\ref{ordgca}) 
are given by  
\bea 
e_1 & = & A\ox I \ox \cdots \ox I \ox I, \nn \\ 
e_2 & = & B \ox I \ox \cdots \ox I \ox I, \nn \\ 
e_3 & = & \mu A^{-1}B \ox A \ox I \ox \cdots \ox I \ox I, \nn \\ 
e_4 & = & \mu A^{-1}B \ox B \ox I \ox \cdots \ox I \ox I, \nn \\ 
e_5 & = & \mu^2 A^{-1}B \ox A^{-1}B \ox A \ox I \ox \cdots \ox I, \nn \\     
e_6 & = & \mu^2 A^{-1}B \ox A^{-1}B \ox B \ox I \ox \cdots \ox I, \nn \\     
& \vdots &  \nn \\ 
e_{2m-1} & = & \mu^{m-1} A^{-1}B \ox A^{-1}B \ox \cdots \ox A^{-1}B \ox A, \nn \\   
e_{2m} & = & \mu^{m-1} A^{-1}B \ox A^{-1}B \ox \cdots \ox A^{-1}B \ox B, \nn \\    
e_{2m+1} & = & \mu^m A^{-1}B \ox A^{-1}B \ox \cdots A^{-1}B \ox A^{-1}B, 
\lb{gcareps} 
\eea 
where $\mu = \o^{(N+1)/2}$ and 
\bea
A & = & 
\lrb\ba{ccccc}
0 & 1 & 0 & \ldots & 0 \\ 
0 &  0 & 1 & \ldots & 0 \\  
\vdots & \vdots   & \vdots & \ddots & \vdots \\ 
0 & 0 & 0 & \ldots & 1\\
1 & 0 & 0 & \ldots & 0 \\
\ea\rrb,   
\lb{A} 
\eea 
\bea 
B & = & 
\lrb\ba{ccccc}
1 & 0 & 0 & \ldots & 0 \\ 
0 & \o & 0 & \ldots & 0 \\
0 & 0 & \o^2 & \ldots & 0 \\ 
\vdots & \vdots   & \vdots & \ddots & \vdots \\ 
0 & 0 & 0 & \ldots &  \o^{N-1} \\
\ea\rrb,    
\lb{B}
\eea 
$N \times N$ unitary matrices, obeying 
\bea 
AB & = & \o BA, \quad A^N\,=\,B^N\,=\,I. 
\lb{abwba}
\eea 
The matrices $A$ and $B$ in (\ref{A}) and (\ref{B}), respectively, provide the only irreducible 
representation for the relation (\ref{abwba}) \cite{W}.  It can also be shown that the GCA 
${\cal C}^{(n)}_N$ defined by (\ref{ordgca}) has only one $N^m$-dimensional irreducible 
representation, as given by (\ref{gcareps}) without $e_{2m+1}$, when $n = 2m$ and there are 
$N$ inequivalent irreducible representations of dimension $N^m$ (differing from (\ref{gcareps}) 
only by multiplications by powers of $\o$) when $n = 2m+1$ (see, {\em e.g.}, \cite{NRRJ1,NRRJ2}).   
This is the generalization of Pauli's theorem for the GCA (\ref{ordgca}).  When $N = 2$ it is 
seen that $A = \s_1$, $B = \s_3$, and the representation (\ref{gcareps}) becomes the representation 
(\ref{careps}) of the Clifford algebra.   

From the structure of the representation (\ref{gcareps}) it is clear that the $\sigma$-operation 
procedure should work in this case also.  Let the $n$-dimensional vector 
$\u{\l} = \{\l_1,\l_2,\ldots,\l_n\}$ be associated with an $\cL$-matrix defined by 
\bea
{\cL}^{(n)} & = & \sum_{j=1}^n \l_j e_j^{(n)} 
\lb{cL}
\eea 
Then, from the commutation relations (\ref{ordgca}) it follows that 
\bea
\lrb {\cL}^{(n)} \rrb^N & = & \lrb \sum_{j=1}^n \l_j^N \rrb I. 
\eea 
Thus, the $N$-th root of $\sum_{j=1}^n \l_j^N$ is given by ${\cL}^{(n)}$ which is linear in 
$\l_j$s.   This fact helps linearize certain $N$-th order partial differential operators using 
the GCA \cite{MN} similar to the way Clifford algebra helps linearize certain second order 
partial differential operators ({\em e.g.}, Dirac's linearization of 
$\h{H}^2 = -\hbar^2c^2\nabla^2 + m^2c^4$ to get his relativistic Hamiltonian 
$\h{H} = -i\hbar c (\a_x{\p}/{\p x} + \a_y{\p}/{\p y} + \a_z{\p}/{\p z}) + mc^2\b$).   Now, it 
can be easily seen \cite{AR} that ${\cL}^{(2m+3)}$ is obtained from ${\cL}^{(2m+1)}$ by the 
$\sigma$-operation: replace $(\l_1,\l_2,\ldots,\l_{2m})$ in ${\cL}^{(2m+1)}$ 
by $(\l_1I,\l_2I,\ldots,\l_{2m}I)$, respectively, where $I$ is the $N$-dimensional identity matrix, 
and $\l_{2m+1}$ by ${\cL}^{(3)}(\l_{2m+1},\l_{2m+2},\l_{2m+3})$.  

From the above it is clear that the matrices $A$ and $B$ in (\ref{A}) and (\ref{B}), respectively, 
obeying the relation (\ref{abwba}), play a central role in the study of GCAs.  If we want to 
have two matrices $A_j$ and $B_j$ obeying 
\bea 
A_jB_j & = & e^{2\pi ij/N}B_jA_j, \quad \mbox{g.c.d}(j,N)\,=\,1, 
\eea 
then, $A_j$ is same as $A$ in (\ref{A}) and $B_j$ is given by $B$ in (\ref{B}) with $\o$ replaced 
by $\o^j$, upto multiplicative factors which are to be determined by the required normalization 
relations like (\ref{norm}).  In the following we shall outline some of the physical applications 
of the matrices $A$ and $B$.  

One approach to study the representation theory of the GCA with ordered $\o$-commutation relations 
(\ref{ordgca}) is to study the vector, or the ordinary, representations of the group 
\bea
{\cal G} & : & \lcb \o^{j_0}e_1^{j_1}e_2^{j_2}\ldots e_n^{j_n} | j_0,j_1,j_2,\ldots j_n 
       = 0,1,2,\ldots N-1 \rcb. 
\eea 
This group has been called a generalized Clifford group (GCG) and the study of its representation 
theory involves interesting number theoretical aspects (\cite{NRRJ1},\cite{NRRJ2}).  Particularly, 
by studying the representations of  the lowest order GCG generated by $A$, $B$ and $\o$ one can show 
that $A$ and $B$ have only one irreducible representation as given by (\ref{A}) and (\ref{B}).   
Study of spin systems defined on a GCG also involves very interesting number theoretical problems 
\cite{JS1}.  Alladi Ramakrishnan and collaborators used the $L$-matrix theory for studying several 
topics like idempotent matrices, special unitary groups arising  in particle physics, algebras 
derived from polynomial conditions, Duffin-Kemmer-Petiau algebra, and para-Fermi algebra (for 
details see \cite{AR}).  They studied essentially the GCA with ordered $\o$-commutation relations 
(\ref{ordgca}).  The more general GCAs (\ref{gca}) were studied later in (\cite{J1}, \cite{RJ}, 
\cite{J2}, \cite{NRRJ1}, \cite{NRRJ2}).   In gauge field theories Wilson operators and 't~Hooft 
operators satisfy commutation relations of the form in (\ref{abwba}) and the corresponding algebra 
is often called the 't~Hooft-Weyl algebra (see, {\em e.g.}, \cite{VPN}).  For the various other 
physical applications of GCAs see, {\em e.g.}, (\cite{K}, \cite{KBJ}).  

\section{Weyl-Schwinger unitary basis for matrix algebra and Alladi Ramakrishnan's matrix 
decomposition theorem}  
\renewcommand\theequation{\thesection.\arabic{equation}}
\setcounter{equation}{0}
Heisenberg's canonical commutation relation between position and momentum operators of a particle, 
the basis of quantum mechanics, is 
\bea
\lsb \h{q} , \h{p} \rsb & = & i\hbar. 
\lb{Hccr} 
\eea 
Weyl \cite{W} wrote it in exponential form as 
\bea
e^{i\eta\h{p}/\hbar} e^{i\xi\h{q}/\hbar} & = & e^{i\xi\eta/\hbar}e^{i\xi\h{q}/\hbar} 
                                                e^{i\eta\h{p}/\hbar}, 
\lb{Wccr} 
\eea 
where the parameters $\xi$ and $\eta$ are real numbers, and studied its representation as the 
large $N$ limit of the relation\,: 
\bea  
AB & = & \o BA, \quad \o\,=\,e^{2\pi i/N}.  
\lb{ABwBA} 
\eea 
Note that the Heisenberg-Weyl commutation relation (\ref{Wccr}) takes the form (\ref{ABwBA}) when 
$\xi\eta/\hbar = 2\pi/N$.   Weyl established that the relation (\ref{ABwBA}), subject to the 
normalization condition 
\bea 
A^N & = & B^N\,=\,I, 
\eea  
has only one irreducible representation as given in (\ref{A}) and (\ref{B}).   Analysing the large 
$N$ limits of $A$ and $B$, he showed that the the relation (\ref{Wccr}), or equivalently the 
Heisenberg commutation relation (\ref{Hccr}), has the unique (upto equivalence) irreducible 
representation given by the Schr\"{o}dinger representation 
\bea
\h{q}\psi(q) & = q\psi(q), 
                 \quad \h{p}\psi(q)\,=\,-i\hbar\fr{d}{dq}\psi(q), 
                 \qquad \mbox{for any}\ \psi(q). 
\eea
This result, or the Stone-von Neumann theorem obtained later by a more rigorous approach, is 
of fundamental importance for physics since it establishes the uniqueness of quantum mechanics.  
Thus, Weyl viewed quantum kinematics as an irreducible Abelian group of unitary ray rotations 
in system space.  

Following the above approach to quantum kinematics Weyl gave his correspondence rule for obtaining 
the quantum operator $\h{f}(\h{q},\h{p})$ for a classical observable $f(q,p)$\,:
\bea
\h{f}(\h{q},\h{p}) & = & 
       \fr{1}{2\pi}\int_{-\infty}^{\infty}\int_{-\infty}^{\infty} d\xi d\eta\,g(\xi,\eta) 
       e^{i(\xi\h{q} + \eta\h{p})}, \nn \\ 
g(\xi,\eta) & = & \fr{1}{2\pi}\int_{-\infty}^{\infty}\int_{-\infty}^{\infty} dq dp\,f(q,p) 
       e^{-i(\xi q + \eta p)}. 
\lb{Wrule} 
\eea 
The fact that the set of $N^2$ linearly independent unitary matrices 
$\{A^kB^l | k,l = 0,1,2,\ldots,(N-1)\}$ forms a basis for the $N \times N$-matrix algebra is 
implicit in this suggestion that any quantum operator corresponding to a classical observable 
can be written as a linear combination of the unitary operators 
$\{e^{i(\xi\h{q} + \eta\h{p})}\}$.  

Schwinger \cite{S} studied in detail the role of the matrices $A$ and $B$ in quantum mechanics 
and hence the set $\{A^kB^l | k,l = 0,1,2,\ldots,(N-1)\}$ is often called Schwinger's unitary 
basis for matrix algebra.  Let us write an $N \times N$ matrix $M$ as  
\bea
M & = & \sum_{k,l = 0}^{N-1} \mu_{kl}A^k B^l.
\lb{M} 
\eea 
From the structure of the matrices $A$ and $B$ it is easily found that 
\bea 
\mbox{Tr} \lsb(A^kB^l)^\dagger(A^mB^n)\rsb & = & N\delta_{km}\delta_{ln}.
\eea 
Hence, 
\bea 
\mu_{kl} & = & 
 \fr{1}{N} \mbox{Tr} \lsb (A^kB^l)^\dagger M \rsb\,=\,\mbox{Tr} \lsb B^{-l}A^{-k}M \rsb. 
\eea 
Alladi Ramakrishnan wrote (\ref{M}) equivalently as 
\bea
M & = & \sum_{k,l = 0}^{N-1} c_{kl} B^k A^l,      
\lb{MAR} 
\eea 
and expressed the coefficients $\{c_{kl}\}$ in a very nice form \cite{AR}\,: 
\bea 
C & = & S^{-1}R, 
\lb{CSR} 
\eea 
\bea 
C & = & \lrb\ba{ccccc}
c_{00} & c_{01} & c_{02} & \ldots & c_{0,N-1} \\ 
c_{10} & c_{11} & c_{12} & \ldots & c_{1,N-1} \\  
c_{20} & c_{21} & c_{22} & \ldots & c_{2,N-1} \\
\vdots & \vdots   & \vdots & \ddots & \vdots \\ 
c_{N-2,0} & c_{N-2,1} & c_{N-2,2} & \ldots & c_{N-2,N-1} \\
c_{N-1,0} & c_{N-1,1} & c_{N-1,2} & \ldots & c_{N-1,N-1} \\
\ea\rrb,    
\lb{C} 
\eea 
\bea
S^{-1} & = & \fr{1}{N} 
\lrb\ba{ccccc}
1 & 1 & 1 & \ldots & 1 \\ 
1 & \o^{-1} & \o^{-2} & \ldots & \o^{-(N-1)}  \\  
1 & \o^{-2} & \o^{-4} & \ldots & \o^{-2(N-1)} \\ 
\vdots & \vdots   & \vdots & \ddots & \vdots \\ 
1 & \o^{-(N-2)} & \o^{-2(N-2)} & \ldots & \o^{-(N-2)(N-1)} \\
1 & \o^{-(N-1)} & \o^{-2(N-1)} & \ldots & \o^{-(N-1)(N-1)} \\
\ea\rrb,   
\lb{S} 
\eea 
\bea
R & = & \lrb\ba{ccccc}
M_{00} & M_{01} & M_{02} & \ldots & M_{0,N-1} \\ 
M_{11} & M_{12} & M_{13} & \ldots & M_{10} \\  
M_{22} & M_{23} & M_{24} & \ldots & M_{21} \\
\vdots & \vdots   & \vdots & \ddots & \vdots \\ 
M_{N-2,N-2} & M_{N-2,N-1} & M_{N-2,0} & \ldots & M_{N-2,N-3}\\
M_{N-1,N-1} & M_{N-1,0} & M_{N-1,1} & \ldots & M_{N-1,N-2} \\
\ea\rrb.  
\lb{R} 
\eea 
Note that $S^{-1}$ is the inverse of the Sylvester, or the finite Fourier transform, matrix.  
He called (\ref{MAR})-(\ref{R}) as a matrix decomposition theorem.   Comparing (\ref{M}) and 
(\ref{MAR}) it is clear that $\mu_{kl} = \o^{-kl}c_{lk}$.  

\section{Finite-dimensional Wigner function}  
\renewcommand\theequation{\thesection.\arabic{equation}}
\setcounter{equation}{0}
Let $N = 2\nu+1$ and choose 
\bea
A & = & 
\lrb\ba{ccccc}
0 & 1 & 0 & \ldots & 0 \\ 
0 &  0 & 1 & \ldots & 0 \\  
\vdots & \vdots   & \vdots & \ddots & \vdots \\ 
0 & 0 & 0 & \ldots & 1\\
1 & 0 & 0 & \ldots & 0 \\
\ea\rrb,   
\eea 
and 
\bea 
B & = & 
\lrb\ba{cccccc}
\o^{-\nu} & 0 & 0 & \ldots & 0 & 0\\ 
0 & \o^{-\nu+1} & 0 & \ldots & 0 & 0 \\
0 & 0 & \o^{-\nu+2} & \ldots & 0 & 0\\ 
\vdots & \vdots   & \vdots & \ddots & \vdots & \vdots \\ 
0 & 0 & 0 & \ldots & \o^{\nu-1} & 0 \\ 
0 & 0 & 0 & \ldots &  0 & \o^{\nu} \\
\ea\rrb.      
\eea 
where $\o = e^{2\pi i/(2\nu+1)}$.  Note that $AB = \o BA$ and $A^{2\nu+1} = B^{2\nu+1} = I$.   
Let us now write a $2\nu+1$-dimensional matrix $M$ as 
\bea 
M & = & \sum_{k,l = -\nu}^\nu v_{kl}\o^{kl/2} B^k A^l, 
\eea 
where 
\bea 
v_{kl} & = & \fr{1}{2\nu+1} \mbox{Tr} \lsb \o^{-kl/2} A^{-l}B^{-k} M \rsb. 
\eea 
If the matrix $M$ is to be Hermitian, {\em i.e.},$M^\dagger = M$, then the condition to be 
satisfied is that $v_{kl}^* = v_{-k,-l}$.  

Let $W= (w_{kl})$, with $k,l = -\nu,-\nu+1,\ldots,\nu-1,\nu$, be a real matrix and define the 
finite-dimensional Fourier transform 
\bea 
v_{\xi\eta} & = & \fr{1}{2\nu+1} \sum_{k,l=-\nu}^\nu w_{kl}\o^{-\xi k - \eta l}. 
\eea 
We have 
\bea 
v_{\xi\eta}^* = v_{-\xi,-\eta}. 
\eea 
Hence, the matrix 
\bea 
H & = & \sum_{\xi,\eta = -\nu}^\nu v_{\xi,\eta}\o^{\xi\eta/2} B^\xi A^\eta \nn \\ 
    & = & \fr{1}{2\nu+1} \sum_{\xi,\eta = -\nu}^\nu 
          \sum_{k,l = -\nu}^\nu w_{kl}\o^{-\xi k -\eta l + (\xi\eta/2)} B^\xi A^\eta 
\lb{H} 
\eea 
is Hermitian.  This property, that to every real matrix $W$ there is associated a unique 
Hermitian matrix $H$, is the  basis of the Weyl correspondence (\ref{Wrule}).   For a given 
Hermitian matrix $H$ the associated real matrix $W$ is obtained from (\ref{H}) as 
\bea
w_{kl} & = & \mbox{Tr} \lsb \o^{\xi k + \eta l - (\xi\eta/2)} A^{-\eta}B^{-\xi} H \rsb. 
\lb{FWF} 
\eea   
In the large $\nu$ limit this provides the converse of the Weyl rule (\ref{Wrule}) for obtaining 
the classical observable corresponding to a quantum operator or the Wigner transform of a quantum 
operator; in particular, the Wigner phase-space quasiprobability distribution function can be 
obtained as the limiting case of (\ref{FWF}) corresponding to the choice of $H$ as the quantum 
density operator \cite{J1}.   Thus, the formula (\ref{FWF}) can be viewed as an expression of 
the finite-dimensional Wigner function corresponding to the case when $H$ is a finite-dimensional 
density matrix.   For more details on finite-dimensional, or discrete, Wigner functions, which are 
of current interest in quantum information theory, see, {\em e.g.}, \cite{Simon}.  

\section{Finite-dimensional quantum canonical transformations}  
\renewcommand\theequation{\thesection.\arabic{equation}}
\setcounter{equation}{0}
As seen above, the relation (\ref{abwba}) has a unique representation for $A$ and $B$ as given 
by (\ref{A}) and (\ref{B}).  Let us take $N$ to be even and make a transformation 
\bea 
A \lra A' & = & \o^{-kl/2} A^kB^l,  \quad 
B \lra B'\,=\,\o^{-mn/2} A^mB^n, 
\lb{fdct} 
\eea 
where $(k,l,m,n)$ can be in general taken to be nonnegative integers in $[0 , N-1]$, and require 
\bea 
A'B' & = & \o B'A', \quad A'^N\,=\,B'^N\,=\,I. 
\eea 
This implies that we should have 
\bea 
kn - lm & = & 1 (\mbox{mod}).N, 
\eea 
and the factors $\o^{-kl/2}$ and $\o^{-mn/2}$ ensure that $A'^N = B'^N = I$.  The uniqueness 
of the representation requires that there should be a definite solution to the equivalence relation 
\bea
SA & = & A'S, \quad SB\,=\,B'S. 
\eea 
Substituting the explicit matrices for $A$ and $B$ from (\ref{A}) and (\ref{B}) it is straightforward 
to solve for $S$.  We get 
\bea 
S_{xy} & = & \o^{-(nx^2-2xy+ky^2)/2m}, \quad x,y\,=\,0,1,2,\ldots,N-1. 
\lb{Sxy} 
\eea 
From the association, following Weyl,  
\bea 
A & \lra & e^{i\eta\h{p}/\hbar}, \quad B\,\lra\,e^{i\xi\h{q}/\hbar}, 
\eea 
it follows that in the limit of $N \lra \infty$ the finite-dimensional transformation (\ref{fdct}) 
becomes the linear canonical transformation of the pair $(\h{q},\h{p})$, 
\bea 
\h{q}' & = & n\h{q} + m\h{p}, \quad \h{p}'\,=\,l\h{q} + k\h{p}. 
\lb{qlct} 
\eea 
By taking the corresponding limit of the matrix $S$ in (\ref{Sxy}) one gets the unitary 
transformation corresponding to the quantum linear canonical transformation (\ref{qlct}) (\cite{J1}, 
\cite{J3}) (for details of the quantum canonical transformations see \cite{MQ}).  

\section{Magnetic Bloch functions}  
\renewcommand\theequation{\thesection.\arabic{equation}}
\setcounter{equation}{0}
For an electron of charge $-e$ and mass $m$ moving in a crystal lattice under the influence of 
an external constant homogeneous magnetic field the stationary state wavefunction corresponding to 
the energy eigenvalue $E$ satisfies the Schr\"{o}dinger equation 
\bea 
\h{\cal H}\psi(\v{r}) & = & E\psi(\v{r}),  \nn \\ 
\h{\cal H} & = & \fr{1}{2m}\lrb \v{\h{p}} + e \v{A} \rrb^2 + V(\v{r}), 
\lb{SchE}
\eea 
where $\v{\h{p}}$ is the momentum operator $-i\hbar\v{\nabla}$, $V(\v{r})$ is the periodic crystal 
potential, and $\v{A} = \fr{1}{2}(\v{B} \times \v{r})$ is the vector potential of the magnetic field 
$\v{B}$.   In the absence of the magnetic field the Hamiltonian is invariant under the group of 
lattice translations and as a consequence  the corresponding wavefunction takes the form of a Bloch 
function\,:  
\bea 
\psi_{\v{B}=0}(\v{r}) & = & \sum_{\v{R}} e^{-i\v{K}\cdot\v{R}} u(\v{r}+\v{R}), 
\lb{Bfn} 
\eea
where $\{\v{R}\}$ is the set of all lattice vectors and $\v{K}$ is a reciprocal lattice vector 
within a Brillouin zone.  This is the basis of the band theory of solids.  In the presence of 
a magnetic field the Hamiltonian $\h{\cal H}$ is not invariant under the lattice translation group.  
Now, the invariance group is the so-called magnetic translation group with its generators given by, 
apart from some phase factors, $\lcb \t_j = e^{i\v{a}_j\cdot(\h{\v{p}}-e\v{A})} | j = 1,2,3 \rcb$ 
where $\v{a}_j$s are the primitive lattice vectors.  These generators obey the algebra\,: 
\bea 
\t_j\t_k & = & e^{-ie\v{B}\cdot\v{a}_j\times\v{a}_k/\hbar}\t_k\t_j, \quad j,k = 1,2,3, 
\eea 
a GCA!   We can obtain the irreducible representations of this algebra in terms of $A$ and $B$ 
matrices.  Once the inequivalent irreducible representations of the magnetic translation group 
are known, using the standard group theoretical techniques we can construct the symmetry-adapted 
basis functions for the Schr\"{o}dinger equation (\ref{SchE}).  This leads to a generalization 
of the Bloch function (\ref{Bfn}), the magnetic Bloch function, given by  
\bea
\psi(\v{r}) & = & \sum_{\v{R}} e^{-i\lsb(\v{K}+\fr{e}{2\hbar}\v{B}\times\v{r})\cdot\v{R} 
                            + \phi(\v{R})\rsb} u(\v{r}+\v{R}), 
\lb{mBfn} 
\eea
where 
\bea
      &       &    \phi(n_1\v{a}_1+n_2\v{a}_2+n_3\v{a}_3) \nn \\ 
      &       &    \quad \quad =\,\fr{e}{2\hbar}\v{B}\cdot(n_1n_2\v{a}_1\times\v{a}_2 + 
                     n_1n_3\v{a}_1\times\v{a}_3 + n_2n_3\v{a}_2\times\v{a}_3).  
\eea   
If the term $\phi(\v{R})$ is dropped from this expression then it reduces to the well known form 
proposed by Peierls (for more details see (\cite{J3}, \cite{NRRJ3}, \cite{NRRJ4}) and references 
therein).  Understanding the dynamics of a Bloch electron in a magnetic field is an important 
problem of condensed matter physics with various  practical applications.  

\section{Finite-dimensional quantum mechanics}  
\renewcommand\theequation{\thesection.\arabic{equation}}
\setcounter{equation}{0} 
Following are the prophetic words of Weyl \cite{W}\,:  {\sl The kinematical structure of a 
physical system is expressed by an irreducible Abelian group of unitary ray rotations in 
system space. ..... If the group is continuous this procedure automatically leads to Heisenberg's 
formulation. .....  Our general principle allows for the possibility that the Abelian rotation 
group is entirely discontinuous, or that it may even be a finite group. ..... But the field of 
discrete groups offers many possibilities which we have not yet been able to realize in Nature; 
perhaps, these holes will be filled by applications to nuclear physics.}  

Keeping in mind the above statement of Weyl and the later work of Schwinger \cite{S}, a 
finite-dimensional quantum mechanics was developed by Santhanam and collaborators.  Following 
Weyl, let us make the association 
\bea 
A & \lra & e^{i\eta\h{p}/\hbar}, \quad B\,\lra\,e^{i\xi\h{q}/\hbar}. 
\eea 
Now if we interpret the finite dimensional matrices $A$ and $B$ as corresponding to 
finite-dimensional momentum and position operators, say, $P$ and $Q$, respectively, with finite 
discrete spectra, then, the corresponding system will have confinement purely as a result of its 
kinematical structure.  The matrices $P$ and $Q$ can be obtained by taking the logarithms of $A$ 
and $B$.  The commutation relation between $P$ and $Q$ was first calculated by Santhanam and 
Tekumalla \cite{ST} (Tekumalla was my senior fellow student at our institute).  Further work by 
Santhanam (\cite{TSS1}-\cite{TSSS}) along these lines resulted in the study of the Hermitian phase 
operator in finite dimensions  as a precursor to the currently well known Pegg-Barnett formalism 
(see, {\em e.g.}, \cite{TMG}).   

Later, we developed a formalism of finite-dimensional quantum mechanics (FDQM)  
(\cite{JSV}-\cite{J4}) in which we studied the solutions of the Schr\"{o}dinger equation with 
finite-dimensional matrix Hamiltonians obtained by replacing the position and momentum operators 
by finite-dimensional matrices $Q$ and $P$.  In \cite{J4} I interpreted quark confinement as a 
kinematic confinement as a consequence of its Weylian finite-dimensional quantum mechanics.   
Recently, dynamics of wave packets has been studied within the formalism of FDQM \cite{BaBe}.  

\section{GCAs and quantum groups}  
\renewcommand\theequation{\thesection.\arabic{equation}}
\setcounter{equation}{0}
Experience of working on GCAs helped me later in my work on quantum groups.  An $n \times n$ 
linear transformation matrix $M$ acting on the noncommutative $n$-dimensional Manin vector 
space and its dual is a member of the quantum group $GL_q(n)$ if its noncommuting elements 
$m_{jk}$ satisfy certain commutation relations.  For example, the elements of a $2 \times 2$ 
quantum matrix belonging to $GL_q(2)$, 
\bea 
M & = & \lrb\ba{cc} m_{11} & m_{12} \\ m_{21} & m_{22} \\ \ea\rrb, 
\eea 
have to satisfy the commutation relations 
\bea 
m_{11}m_{12} & = & q^{-1}m_{12}m_{11}, \quad m_{11}m_{21}\,=\,q^{-1}m_{21}m_{11}, \nn \\
m_{12}m_{22} & = & q^{-1}m_{22}m_{12}, \quad m_{21}m_{22}\,=\,q^{-1}m_{22}m_{21}, \nn \\ 
m_{12}m_{21} & = & m_{21}m_{12}, \quad 
   m_{11}m_{22} - m_{22}m_{11}\,=\,(q^{-1} - q)m_{12}m_{21}.   
\eea 
Some of these relations are already GCA-like, or Heisenberg-Weyl-like.  It was shown in 
(\cite{F},\cite{We}) that, in general, all the commutation relations of $GL_q(n)$ can be 
formulated in a similar form and hence the representations of these elements can be found utilising 
the representation theory of the Heisenberg-Weyl relations. Extending these ideas further, we 
developed in \cite{CJ1} a systematic scheme for constructing the finite and infinite dimensional 
representations of the elements of the quantum matrices of $GL_q(n)$, where $q$ is a primitive 
root of unity, and discussed the explicit results for $GL_q(2)$, $GL_q(3)$ and $GL_q(4)$.  In 
this work we essentially used the product transformation technique (\cite{J1},\cite{RJ}) 
developed in the context of representation theory of GCAs.  In \cite{CJ2} we extended this formalism 
to the two-parameter quantum group $GL_{p,q}(2)$ and the two-parameter quantum supergroup 
$GL_{p,q}(1|1)$.   

\section{Conclusion}  
\renewcommand\theequation{\thesection.\arabic{equation}}
\setcounter{equation}{0}
To summarize, I have reviewed here some aspects of GCAs and their physical applications, mostly 
related to my own work.  I learnt about it in the school of Alladi Ramakrishnan and it has 
been useful to me throughout my academic career so far.  I would like to conclude with the 
following remark on GCAs by Alladi Ramakrishnan \cite{AR2}\,: 
\begin{quote}
{\em The structure is too fundamental to be unnoticed, too consistent to be ignored, and much 
too pretty to be without consequence.}  
\end{quote} 
\vspace{1cm} 
\noindent 
{\em Acknowledgement}\,: {\sf I dedicate this article, with gratitude, to the memory of my 
teacher Professor Alladi Ramakrishnan under whose guidance I started my scientific career at 
MATSCIENCE, The Institute of Mathematical Sciences, Chennai. }

\end{document}